# Spin measurement using cycling transitions of a two-electron quantum dot molecule

Y. L. Delley,[1] M. Kroner,[1] S. Faelt,[1,2] W. Wegscheider,[2] and A. İmamoğlu[1]

[1]*Institute for Quantum Electronics, ETH Zürich, CH-8093 Zurich, Switzerland.*
[2]*Laboratory for Solid State Physics, ETH Zürich, CH-8093 Zurich, Switzerland.*
(Dated: September 14, 2015)

Two-electron charged self-assembled quantum dot molecules exhibit a decoherence-avoiding singlet-triplet qubit subspace and an efficient spin-photon interface. Here, we demonstrate that the cycling transitions originating from auxiliary ground states in the same system allow for an efficient optical read-out of a singlet-triplet qubit. By implementing a spin-selective state transfer to the auxiliary state using a resonant laser field, we observe an improvement approaching two orders of magnitude in fidelity as compared to spin measurement by light scattering directly from the qubit states. Embedding the quantum dot molecule inside a low quality-factor micro-cavity structure should enable single-shot qubit read-out.

PACS numbers: 03.67.Lx, 73.21.La, 42.50.-p

Spins in optically active quantum dots (QD) exhibit relatively short $T_2^*$ coherence times. Despite this strong limitation, QDs stand out among solid-state qubit systems for their excellent optical properties that render them promising for quantum communication tasks relying on a quantum interface between stationary (spin) and flying (photonic) qubits. Recent experiments have used this favorable feature to demonstrate coherent all-optical spin manipulation [1], spin-photon entanglement [2–4], teleportation from a photonic to a spin qubit [5] and heralded distant spin entanglement using QDs in Voigt geometry [6]. Further progress in these experiments is limited by the lack of efficient spin measurement.

With this background, singlet-triplet ($|S\rangle$–$|T_0\rangle$) states in optically active quantum dot molecules (QDM) in Faraday geometry emerge as promising candidates for quantum information processing since (i) they exhibit decoherence-avoiding clock-transitions that are insensitive to fluctuations in both electric and magnetic fields, (ii) the qubit states exhibit equal coupling strength to common optically excited trion states that is essential for maximal spin-photon entanglement, and (iii) the spin polarized triplet states ($|T_+\rangle$ and $|T_-\rangle$) of the ground-state manifold exhibit cycling optical transitions. In this Letter, we show how the latter feature could be used to enhance spin measurement efficiency by almost two orders of magnitude.

*S–T₀ qubits in QDMs* Our experiment is carried out on a single InGaAs self-assembled QDM, consisting of two QDs separated by a 9 nm GaAs tunneling barrier. A semi-transparent metallic top gate and a back $n$-doped layer form a Schottky diode, which is used to control the charge state of the QDM. Thanks to engineered confinement energies in the two QDs, the QDM can be brought into the (1,1)-regime [7, 8], where each QD is charged with a single electron. In this regime, the singlet state ($|S\rangle$) is split from the triplet states ($|T_0\rangle$, $|T_+\rangle$ and $|T_-\rangle$) by the exchange splitting, which is gate-voltage tunable and has a minimum value of $E_{\rm ST} \approx 318\,\mu$eV in our structure. The triplets are split by $E_B \approx 31\,\mu$eV from each other by an external magnetic field of 1 T that is applied along the growth direction (Faraday geometry). The relevant level scheme and the optical transitions are outlined in Fig. 1. Under these conditions, $|S\rangle$ and $|T_0\rangle$ can be compared to atomic clock transitions that are insensitive to both electric and magnetic field fluctuations [9]. Coupling to the common optically excited state $|R_+\rangle$ (with equal oscillator strength) allows for manipulation of the qubit [7, 8]. However, this coupling makes measurement of the qubit by detection of spin-dependent resonance fluorescence (RF) very difficult, as the spin information is quickly lost due to spin-pumping after scattering only two photons on average. Our measurement protocol avoids this by spin-selectively transferring the population of one qubit state to one of the spin-polarized states $|T_+\rangle$ and $|T_-\rangle$. While the latter are sensitive to Overhauser field fluctuations and thus are less ideal for the storage of quantum information, they have only one strongly-allowed optical transition to $|R_{++}\rangle$ and $|R_{--}\rangle$ respectively. As a consequence, $|T_+\rangle \leftrightarrow |R_{++}\rangle$ and $|T_-\rangle \leftrightarrow |R_{--}\rangle$ are *cycling transitions*. Population in $|T_+\rangle$ or $|T_-\rangle$ is much easier to detect, as the cycling property allows many photons to be scattered before spin-pumping destroys the information. "Diagonal" transitions between the cycling subspaces $\{|T_-\rangle, |R_{--}\rangle\} / \{|T_+\rangle, |R_{++}\rangle\}$ and the "lambda-system" subspace $\{|S\rangle, |T_0\rangle, |R_+\rangle\}$ hosting the qubit are only weakly allowed due to heavy-light hole mixing [10]. Their oscillator strength is reduced by a factor $\sim$200 compared to $|T_-\rangle \leftrightarrow |R_{--}\rangle$ and $|T_+\rangle \leftrightarrow |R_{++}\rangle$. The lifetime of the neutral exciton in the top dot is 0.4 ns; we only address the optical transitions in this dot.

*Experimental setup* The sample is held in a liquid helium bath cryostat at 4.2 K. A solid immersion lens is placed on the sample to enhance the extraction efficiency into a NA = 0.68 objective. Three tunable diode lasers are sent through electro-optic modulators (EOM) used as fast shutters and combined to produce the excitation pulses. The EOM contrast ratios are continuously monitored and the biases automatically readjusted whenever necessary to keep the contrast above 1:100. A confocal microscope is used to focus the laser pulses onto the QDM as well as to collect the RF photons from the QDM. A cross-polarization technique [11, 12] is used to suppress the reflected laser background to below $10^{-6}$. Linear polarization is used such that all transitions couple to excitation and detection equally well. The collected photons are detected on a Si avalanche photo-diode and their arrival times are registered with sub-ns resolution in a histogram synchronized with the pulse pattern used to excite the QDM. The overall detection efficiency of our setup is about $\sim 2.5 \times 10^{-4}$. It is limited mainly due to the mode mismatch between the QDM emission and our collection optics.

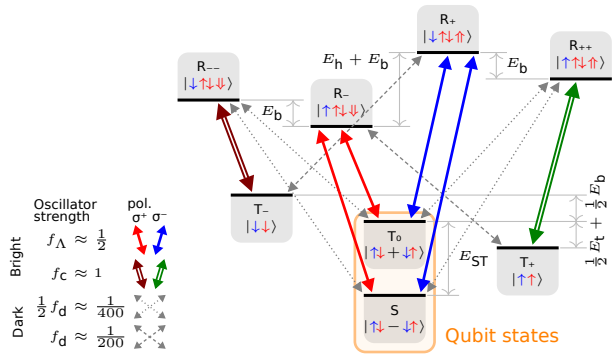

FIG. 1: Spin states and optically excited states (involving an additional electron-hole pair in the top dot) in (1,1) regime. In the state kets, single arrows denote electron spins, double arrows denote hole spins. Spins denoted by a red (blue) arrow are mainly located in the top (bottom) dot. The effective qubit is made up of the states $|S\rangle$ and $|T_0\rangle$. Full lines denote circularly polarized and strongly allowed transitions, dashed and dotted lines denote weakly allowed transitions. $E_t$ and $E_b$ indicate the Zeeman energies for electrons in the top and bottom dot respectively. $E_h$ indicates the Zeeman energy for holes in the top dot.

*Pulse sequence* The pulse sequence used to verify our spin-measurement protocol is outlined in Fig. 2. Our realization of the protocol measures the $|T_0\rangle$ population by transferring it to the ancillary $|T_+\rangle$ state. This is achieved using a 100 ns pulse of laser 2, set resonant with the $|T_0\rangle \leftrightarrow |R_{++}\rangle$ transition (Fig. 2(a): *transfer*). The laser is set to the maximal power of $\sim 190$ nW, corresponding to about $7.2\times$ above saturation of the cycling transitions (relative oscillator strength $f_c \sim 1$), but far from saturation of the diagonal transition. During this pulse, population in $|T_0\rangle$ will eventually get excited to $|R_{++}\rangle$ from where it decays to $|T_+\rangle$ with very high probability. Once in $|T_+\rangle$, the population can be effectively measured by detection of the RF of laser 3 tuned to $|T_+\rangle \leftrightarrow |R_{++}\rangle$ (Fig. 2(a): *measurement*). At 24 nW, the transition is close to saturation.

To quantify the effectiveness of our spin-measurement protocol, we either prepare the qubit in $|T_0\rangle$ to determine the detection efficiency of the measurement scheme, or in $|S\rangle$ to derive an upper limit on the background signal or "false positives". Initialization into $|S\rangle$ is achieved by enabling both laser 2 and laser 3 during 1 µs, spin-pumping population from all triplet states into $|S\rangle$ (Fig. 2(a): *initialization*). In order to prevent coherent population trapping [9, 13] in a superposition of the triplet states, the lasers are rapidly alternated rather than being switched on simultaneously. Due to the degeneracies among the transitions involving the triplets, deterministic one-step spin-pumping into $|T_0\rangle$ is not possible. Instead, to prepare $|T_0\rangle$, we first initialize into $|S\rangle$ and then apply a 20 ns pulse of laser 1, resonant with $|S\rangle \leftrightarrow |R_+\rangle$ (Fig. 2(a): *preparation*). The only other strongly allowed transition from $|R_+\rangle$ leads to $|T_0\rangle$, thus fast spin-pumping will quickly prepare $|T_0\rangle$ with high fidelity. The power of laser 1 is set to 48 nW, close to saturation power of the addressed transition.

The results for our spin-measurement protocol are compared with direct measurement based on spin-selective RF scattering on $|T_0\rangle \leftrightarrow |R_+\rangle$, using a shorter 100 ns pulse of laser 3. The energies of $|T_+\rangle \leftrightarrow |R_{++}\rangle$ and $|T_0\rangle \leftrightarrow |R_+\rangle$ differ only by the difference in g-factor of the two QDs, which at 1 T amounts to less than 1 µeV (refer to Fig. 1). Therefore, laser 3 can drive both transitions nearly resonantly using a single wavelength.

In order to ensure equal experimental conditions, all four combinations (initialization into either spin-state, cycling or direct measurement) are repeated once per 20.4 µs in an interleaved fashion.

*Spin measurement results* The histograms of detected RF photons are shown in Fig. 2(b) for our cycling measurement protocol and Fig. 2(c) for direct measurement. The background signal level due to residual reflected laser light is measured continuously. This is achieved by applying a square modulation to the gate at 731 Hz, periodically ejecting all electrons from the QDM and thus disabling RF scattering.

When restricting the detection to the first 1.3 µs of the measurement pulse, we find a $(3.76 \pm 0.03)\%$ probability to correctly detect the $|T_0\rangle$ state, and a $(0.73 \pm 0.02)\%$ probability for a false positive detection. Without the spin transfer (direct detection), in the first 25 ns of measurement, the probability to detect $|T_0\rangle$ is $(4.8 \pm 0.4) \times 10^{-4}$ with $(1.4 \pm 0.2) \times 10^{-4}$ false positive detection. Even when utilizing the cycling transition, we are still in the spin *heralding* regime; the probability to detect a photon is low, such that the lack of any detection event does not reveal much information about the spin. In that regime, the probability to detect a second photon during the same spin measurement can be neglected, and the heralding probability is proportional to the integrated photon detection rate.

Let us now discuss the length of the spin measurement detection window. Due to spin-pumping, both the signal as well as the false positives will exponentially decay until a spin-independent background level is reached [shown in Fig. 3(a)]. Therefore, while the detection probability of spin-dependent photons saturates, the spin-independent background does not. A compromise has to be found between maximizing the heralding rate and it's "accuracy"; the more false positives are detected, the less likely a photon detection event correctly identifies $|T_0\rangle$.

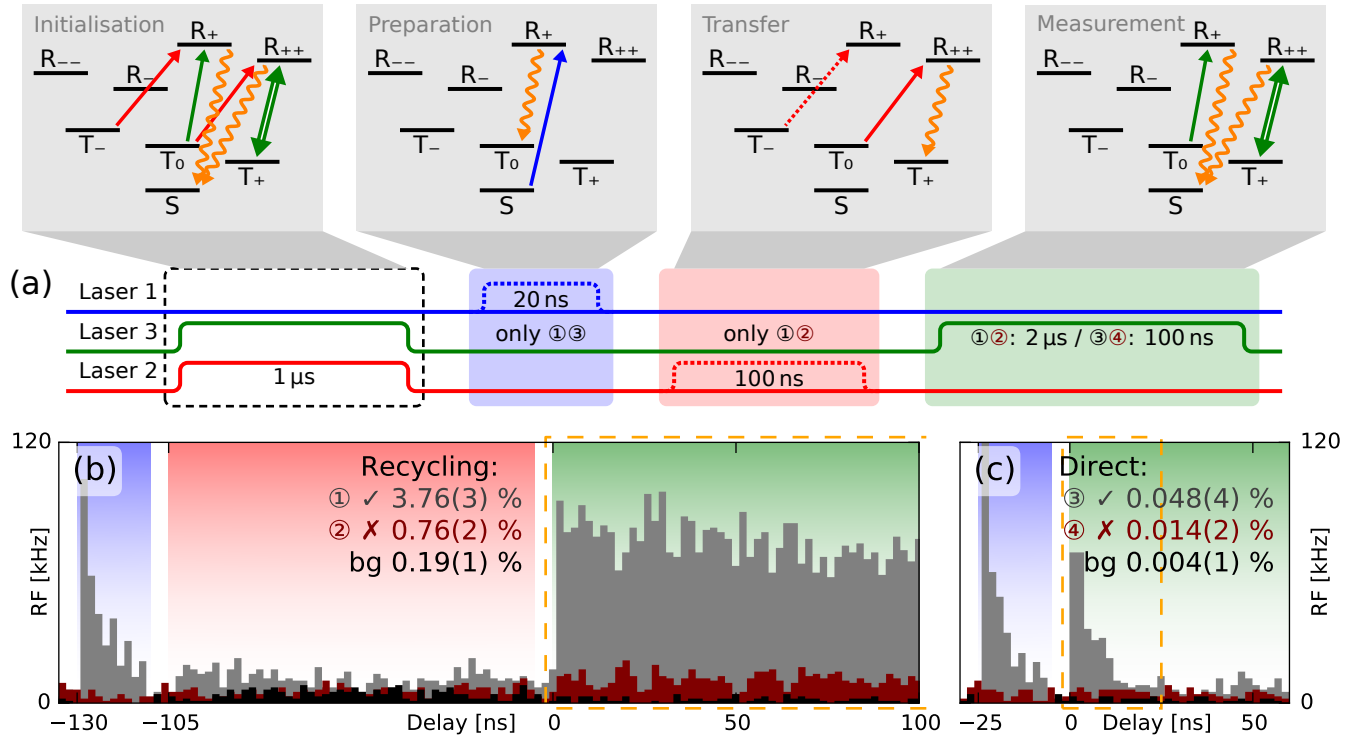

FIG. 2: (a) The pulse sequences used in the experiment. For each pulse, sketches indicate the transitions of the level scheme that are resonantly driven. Wavy orange lines indicate the most relevant spontaneous emission paths that lead to spin-pumping. The full sequence alternates between the four variations ①②③④ in an interleaved fashion, differing in which pulses are used and which are omitted: When the *preparation* pulse is used, the qubit is prepared in $|T_0\rangle$, otherwise it is left in $|S\rangle$. If the *transfer* pulse is used, a cycling measurement is performed, otherwise a direct measurement is performed. (b) Gray (dark red) histogram shows time-resolved RF count rate of the cycling transition spin measurement when the qubit is prepared in $|T_0\rangle$ ($|S\rangle$) by presence (absence) of the preparation pulse from laser 1, corresponding to correct (false) detection of $|T_0\rangle$. The binning size is 2 ns and the histogram was integrated for a total of $5 \times 10^5$ repetitions. Blue, red and green shaded backgrounds indicate time of preparation, transfer and measurement pulses respectively. Black histogram shows residual laser background. Laser background is subtracted from histogram for presentation purposes during the preparation pulse only (blue shaded area), as there it is particularly high and hampering comparability. The background count rate from the preparation laser is about 25 kHz. The values given are the detection probabilities integrated over 1.3 µs. (c) as in (b), but for direct spin measurement, i.e. with the transfer pulse omitted. The orange dashed rectangle indicates the integration window of 25 ns.

The figure of merit commonly used for quantum measurements is the fidelity. It is the average of the probabilities to correctly detect a qubit in either state. For a heralding measurement, the fidelity is therefore just the difference between the detection probability and the false positives plus 50%. Obviously, any spin-independent background signal does not enter the fidelity. Indeed, as shown in Fig. 3(b), there is no penalty for prolonging the measurement time when only spin-independent background is detected. The window length of 1.3 µs for the cycling measurement and 25 ns for the direct measurement were instead selected based on a measure for the "usefulness" of the obtained data: the *mutual information* as defined e.g. in [14] between the measurement result and the prepared spin.

Our protocol improves the heralding rate over direct detection by about a factor of 80, slightly short of the two orders of magnitude to be expected from the branching ratios of $|T_+\rangle \leftrightarrow |R_{++}\rangle$ and $|T_0\rangle \leftrightarrow |R_+\rangle$. We attribute the discrepancy to imperfect population transfer from $|T_0\rangle$ to $|T_+\rangle$. Indeed, off-resonant excitation of other exciton levels puts a limit on the population transfer efficiency. At 1 T, the diagonal transition $|T_0\rangle \leftrightarrow |R_{++}\rangle$ is about 10 (natural) line widths detuned from the vertical transition $|T_0\rangle \leftrightarrow |R_+\rangle$. As the oscillator strength of that transition is stronger by a factor $\frac{f_\Lambda}{f_d} \approx 200$ however, the ratio of off-resonant $|R_+\rangle$ excitation to the desired $|R_{++}\rangle$ excitation is significant. Higher magnetic fields would reduce population loss due to off-resonant excitation, but if would complicate the verification due to the splitting of the remaining nearly-resonant transitions: Direct measurement and cycling measurement would require separate drive lasers, and to prevent population shelving in $|T_-\rangle$, a second extra laser would be necessary. Furthermore, dynamic polarization effects of the nuclear spins become more pronounced at higher fields [15], complicating the comparison of different parameter sets. Similar reasoning holds true for $|T_0\rangle \leftrightarrow |R_-\rangle$, albeit at a somewhat lower rate. A simple estimate based on rate equations predicts 30 % of the $|T_0\rangle$ population will eventually drop to $|S\rangle$ via off-resonant excitation of $|R_+\rangle$ (details in the appendix in section ).

Further insight about the transfer efficiency can be gained by analyzing the influence of the transfer pulse parameters. As long as the diagonal transition is driven far below saturation, the transfer efficiency is only dependent on the integrated pulse power. As with the detection window length, increasing the transfer pulse area comes at the cost of an increased false positive detection rate. Figure 3(c) and (d) shows the results we obtained for different transfer pulse lengths and powers. A striking point is the fact that the false positives rate does not saturate, while the $|T_0\rangle$– $|T_+\rangle$ transfer efficiency does. This suggests that the false positives are primarily due to population excited out of $|S\rangle$ *during the transfer pulse*, instead of residual $|T_0\rangle$ or $|T_+\rangle$ population present before the transfer pulse. The data agrees quite well with a simple model for the spin-transfer from $|T_0\rangle$ to $|T_+\rangle$ based on rate equations. For the case of the qubit prepared in the $|T_0\rangle$ state, we use our previous estimates of the branching ratios, saturation powers and collection efficiency to arrive at a model with no free parameters. Details can be found in the appendix. What cannot be explained by such a model is the lower efficiency seen when using 200 ns pulses compared to shorter pulses. We assume this to be a consequence of slightly different experimental conditions in effect when that dataset was taken. To model the false positives, we include a direct population transfer from $|S\rangle$ to $|T_+\rangle$ at a rate proportional to the transfer from $|T_0\rangle$ to $|T_+\rangle$. The data of figure 3(d) is fitted using a proportionality constant of 5 %. We attribute the source of that population transfer to residual optical coupling of $|S\rangle$ to the excitons due to incomplete suppression of laser 1 by the EOM.

*Outlook* In summary, we have established that cycling transitions in QDMs in the (1,1) regime allow for a significant improvement in qubit measurement efficiency. While the overall efficiency we have demonstrated is modest due to our low photon collection efficiency of $2.5 \times 10^{-4}$, achieving single-shot read-out of $|S\rangle$–$|T_0\rangle$ qubits is within reach. To this end, we note that embedding QDs in a low Q cavity has been shown to improve the mode-matching and to yield an overall detection efficiency exceeding 1 % [6]. The protocol suffers from the difficulty to address individual transitions, impacting the fidelity of the population transfer. This issue can be reduced by the use of higher magnetic fields, possibly allowing for up to 50 % higher detection efficiency. However, this comes at the cost of increased nuclear spin effects and a more complicated spin initialization. We emphasize that even with the improvements in spin measurement efficiency reported here, realization and verification of heralded entanglement of distant $|S\rangle$–$|T_0\rangle$ qubits is feasible [6].

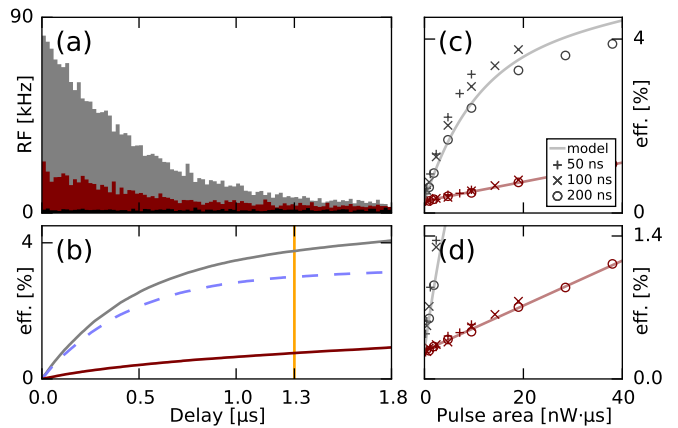

FIG. 3: (a) Grey histogram: Full 1.8 µs histogram of cycling measurement binned at 20 ns bin sizes. Dark red histogram: False-positives during cycling measurement. Black histogram: Laser background. Delay is time since the start of the measurement pulse. (b) Grey (dark red) line: Total integrated $|T_0\rangle$ detection efficiency (false positives rate) vs. width of detection window. Blue dashed line: The difference between correct detection and false positive detection, equivalent to the spin detection fidelity minus 50 %. The orange vertical line indicates detection window of 1.3 µs which maximizes mutual information between measurement and spin. (c) $|T_0\rangle$ detection efficiency vs. transfer pulse area, measured for three different transfer pulse lengths. Full line indicates expected efficiency based on a simple rate equation model using the parameters given in the main text (see Appendix). (d) Zoom-in of (c) around false positive detection.

### ACKNOWLEDGMENTS

This work is supported by NCCR Quantum Science and Technology (NCCR QSIT), research instrument of the Swiss National Science Foundation (SNSF) and by Swiss NSF under Grant No. 200021-140818.

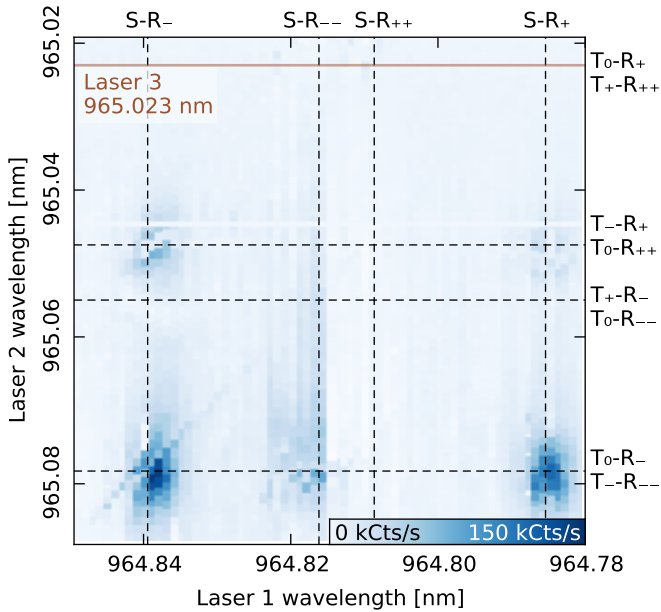

FIG. 4: Three-laser RF spectroscopy, color indicates RF count rate detected from QDM. The wavelength of laser 1 is stepped, laser 2 scanned and laser 3 kept fixed. Dashed lines indicate resonances with QDM transitions.

|  | ES: | | | |
|---|---|---|---|---|
| GS: | $\|R_{--}\rangle$ | $\|R_{-}\rangle$ | $\|R_{+}\rangle$ | $\|R_{++}\rangle$ |
| $\|T_{-}\rangle$ | $4.2 \times 10^{-3}$ | 0.0 | $1.0 \times 10^{-2}$ | 0.0 |
| $\|T_{0}\rangle$ | $2.2 \times 10^{-4}$ | $2.1 \times 10^{-3}$ | $3.3 \times 10^{-3}$ | $5.2 \times 10^{-3}$ |
| $\|T_{+}\rangle$ | 0.0 | $4.3 \times 10^{-4}$ | 0.0 | $6.6 \times 10^{-3}$ |
| $\|S\rangle$ | $1.5 \times 10^{-7}$ | $4.4 \times 10^{-5}$ | $2.8 \times 10^{-5}$ | $1.4 \times 10^{-7}$ |

TABLE I: Calculated excitation level: Ratio between excited state (ES) population and total population in excited state and ground state (GS), induced by laser 3 (*transfer laser*).

# APPENDIX

### Identification of dark transition energies

The presence of four non-degenerate metastable ground states makes high-resolution spectroscopy of weak transitions non-trivial. Fast spin-pumping renders most optical signatures weak. It can only be overcome by addressing all ground states simultaneously.

In order to identify the transition energies, our three lasers are enabled simultaneously and continuously (figure 4). Laser 3 is fixed at 965.023 nm, resonant with $|T_0\rangle \leftrightarrow |R_+\rangle$ and $|T_+\rangle \leftrightarrow |R_{++}\rangle$, which can be found by extrapolation from the edges of the (1,1) charge plateau. Laser 1 is stepped over the transitions involving the singlet, while laser 2 is independently swept over the transitions involving the triplets. Significant RF is only registered when laser 2 is resonant with either $|T_-\rangle \leftrightarrow |R_+\rangle$ or $|T_-\rangle \leftrightarrow |R_{--}\rangle$, and at the same time laser 1 is resonant with $|S\rangle \leftrightarrow |R_-\rangle$ or $|S\rangle \leftrightarrow |R_+\rangle$. In these cases, all four ground-state are driven by a laser field, and spin-pumping is inhibited. With the exception of driving $|S\rangle \leftrightarrow |R_{--}\rangle$ and simultaneously $|T_-\rangle \leftrightarrow |R_{--}\rangle$, coupling $|S\rangle$ using a diagonal transition does not recover RF, most likely due to dynamic nuclear spin polarization effects [15].

### Rate equations model of spin pumping

Since all time-scales of the laser pulses used in the experiment are slower than the spontaneous emission lifetime, the QDM will be quickly driven to steady-state with the incident laser power. Starting from an eigenstate of the system, as long as optically induced population transfer is slow, no significant coherences between ground states build up. Consequentially, the dynamics of our system can be described using rate equations at all times.

We apply the two level atom steady-state result [16] to each transition individually: If a given transition with (relative) oscillator strength $f$ is driven with a coherent field of power $\Omega^2$ and detuning $\Delta^2$, the transition ground state population $p_g$ and the optically excited state population $p_e$ will stay in the fixed ratio

$$\frac{p_e}{p_e + p_g} = \frac{1}{2} \frac{fs}{1 + fs + \frac{\Delta^2}{\gamma^2}}, \quad (1)$$

where $fs = f\frac{\Omega^2}{\gamma^2}$ is the saturation parameter of the transition. Table I displays the ratio $p_e/(p_g + p_e)$ for the parameters of our transfer pulse; $s = 7.2$. Indeed, all the excited state populations stay small, consistent with our assumptions. While the resonant transitions $|T_0\rangle \leftrightarrow |R_{++}\rangle$ and $|T_-\rangle \leftrightarrow |R_+\rangle$ indeed get the highest excitation level, both vertical transitions $|T_0\rangle \leftrightarrow |R_+\rangle$ and $|T_0\rangle \leftrightarrow |R_-\rangle$ are driven to levels within the same order of magnitude. Also, the reverse diagonals $|T_+\rangle \leftrightarrow |R_-\rangle$ and $|T_0\rangle \leftrightarrow |R_{--}\rangle$ are suppressed by little more than one order of magnitude, as they are not too far from resonance.

Spontaneous emission via other transitions will lead to spin-pumping. The rate at which population is transferred is the product of the spontaneous emission rate $\gamma$, the relative oscillator strength $f'$ and the excited state population $p_e$:

$$\dot{p}'_g = \gamma f' p_e. \quad (2)$$

Since $p_e/p_g \ll 1$ we set $\dot{p}_g = -\dot{p}'_g$. Combining equations (1) and (2), we get linear coupled differential equations between the ground state populations. Furthermore, under our conditions, expression (1) is approximately linear in the saturation level $s$. Then, the amount of population that is spin-pumped only depends on the pulse area integral.

| Origin: | Total rate ($\mu s^{-1}$) | Destination: $|T_-\rangle$ (%) | $|T_0\rangle$ (%) | $|T_+\rangle$ (%) | $|S\rangle$ (%) |
|---|---|---|---|---|---|
| $|T_-\rangle$ | 25.9 |  | 50 | 0 | 50 |
| $|T_0\rangle$ | 20.5 | 3 |  | 64 | 33 |
| $|T_+\rangle$ | 1.1 | 0 | 50 |  | 50 |
| $|S\rangle$ | 0.6 | 0 | 100 | 0 |  |

TABLE II: Calculated optical spin pumping rates between (1,1) states of QDM induced by laser 3 (*transfer laser*).

The total rate of spin-pumping out of each state and the relative distribution of transfer destinations is listed in table II, for the parameters of our transfer pulse. One third of the $|T_0\rangle$ population is lost to $|S\rangle$ instead of being transferred to $|T_+\rangle$. This is due to the large off-resonant excitation of $|R_+\rangle$ and $|R_-\rangle$. Furthermore, the reverse population transfer back from $|T_+\rangle$ is non-vanishing. With longer and/or higher intensity transfer pulses, this will eventually lead to a reduced transfer efficiency. Finally, we find a an even weaker rate of off-resonant excitation of $|S\rangle$ population to $|T_0\rangle$, mainly via $|R_-\rangle$. Together with estimates for the collection efficiency and the saturation power, the rates listed above build the model used in figure 3(c) of the main text. However, the off-resonant excitation rate out of $|S\rangle$ is not enough to explain the false positives we measure in our experiment. Instead, already a small amount of residual light from laser 1 driving $|S\rangle \leftrightarrow |R_+\rangle$ would excite $|S\rangle$ population faster than laser 3. Such residual light would be caused by the finite suppression level of the EOMs. For Fig. 3(d), a direct $|S\rangle$ to $|T_+\rangle$ excitation channel at a rate of 5 % of the rate from $|T_0\rangle$ to $|T_+\rangle$ was introduced.